# Parametric metasurfaces for amplified up-conversion of electromagnetic waves

Fedor Kovalev* and Ilya Shadrivov

ARC Centre of Excellence for Transformative Meta-Optical Systems (TMOS), Research School of Physics, The Australian National University, Canberra ACT 2601, Australia
*Contact author: fedor.kovalev@anu.edu.au

Spatiotemporal metasurfaces, characterized by dynamic variations in both space and time, enable functionalities unattainable with passive metasurfaces. In this study, we propose a novel concept of parametric metasurfaces capable of performing frequency up-conversion and amplification of free-space electromagnetic waves. This functionality is realized through the modulation of variable capacitances embedded in specially designed dual split-ring resonators. We explore various combinations of the pump and incident wave frequencies and investigate different operational regimes of the proposed parametric system. Our theoretical analysis reveals high up-conversion efficiency, with amplification levels exceeding 20 dB through a cascaded process, surpassing the limits set by the Manley-Rowe relations. Moreover, precise control over the phase of the up-converted waves is achieved by tuning the pump phase, paving the way for advanced manipulation of electromagnetic waves. This approach is applicable in the microwave and subterahertz ranges, with potential scalability to higher frequencies through ultrafast modulation techniques.

Over the past three decades, significant progress has been made in the development of metamaterials and metasurfaces, which are designed to exhibit a range of unique properties. While static metasurfaces have already found practical applications[1,2], particularly in the miniaturization of optical elements[3,4], the current focus has shifted towards reconfigurable metasurfaces and approaches enabling the dynamic change of their properties[5–9]. This dynamic aspect represents an active area of research, holding the promise of being the next breakthrough in the evolution of metamaterials[10–15].

Metasurfaces, whose characteristics vary over time, are commonly referred to as time-varying[11,14] or parametric[16,17]. They offer opportunities to overcome the fundamental limitations of passive electromagnetism[18] and realize effects previously unattainable with conventional metamaterials, such as magnetic field-free nonreciprocity[19,20]. Recently, time-varying metasurfaces provided a platform for realizing time crystals[21,22], a temporal analogue of the Young's double-slit diffraction experiment[23], beam steering[24,25], frequency conversion[26–32] and electromagnetic wave amplification[33–38].

There is a growing interest in leveraging time-varying media for combined amplification and up-conversion[39,40]. However, research on parametric amplification of free-space electromagnetic waves combined with frequency up-conversion using time-varying metasurfaces remains limited. Only a few theoretical works[34,38] have predicted the feasibility of achieving parametric up-conversion with amplification, but the reported frequency conversion ratios and gains in these transmission line-based systems are relatively low. In these works, the conversion efficiency was enhanced by utilizing reflective designs, making them bulky and less practical for integration into compact systems. For a metasurface without the reflector, the reported frequency conversion efficiency was only 23%.

In our work, we introduce a dual-resonance metasurface which up-converts and amplifies incident electromagnetic waves. Each meta-atom of the proposed parametric metasurface comprises two connected split-ring resonators (SRR) with an embedded varactor diode that is electrically modulated at the pump frequency. One resonator is dedicated to resonantly exciting currents by the incident wave at the signal frequency, while the second SRR resonates at the sum-frequency. We optimize this structure and predict that our parametric metasurface achieves significant up-conversion efficiency with amplification exceeding 20 dB by utilizing a cascaded process, surpassing the limitations imposed by the Manley-Rowe relations. Moreover, by tuning the phase of the modulation, we gain precise control over the phase of the up-converted waves, offering additional possibilities for advanced wave manipulation. Our findings are particularly relevant for applications in the microwave and terahertz frequency ranges, with the potential for extending to higher frequencies through ultrafast modulation techniques[41,42].

Dual-resonance parametric metasurfaces enable the amplification and conversion of incident radiation by employing auxiliary oscillations excited at combination frequencies $f_{comb} = mf_p \pm nf_s$, where $f_s$ is the incident wave (signal) frequency, $f_p$ is the pump frequency, $m$ and $n$ are integers. This study introduces a parametric metasurface with variable capacitance designed to amplify incident waves and up-convert them to the combination frequencies $f_{comb} = mf_p \pm nf_s$, as illustrated in Fig. 1.

Figure 1 shows the meta-atom concept of the proposed dual-resonance parametric metasurface comprising two connected split-ring resonators with an embedded varactor diode that is electrically modulated at the pump frequency $f_p$. The developed parametric system utilizes the nonlinear capacitance of varactor diodes to mix currents excited by the incident wave and the pump source. One resonator serves to effectively excite currents by incident electromagnetic waves, while another resonator is designed for resonant radiation at one of combination frequencies $f_{comb} = mf_p \pm nf_s$.

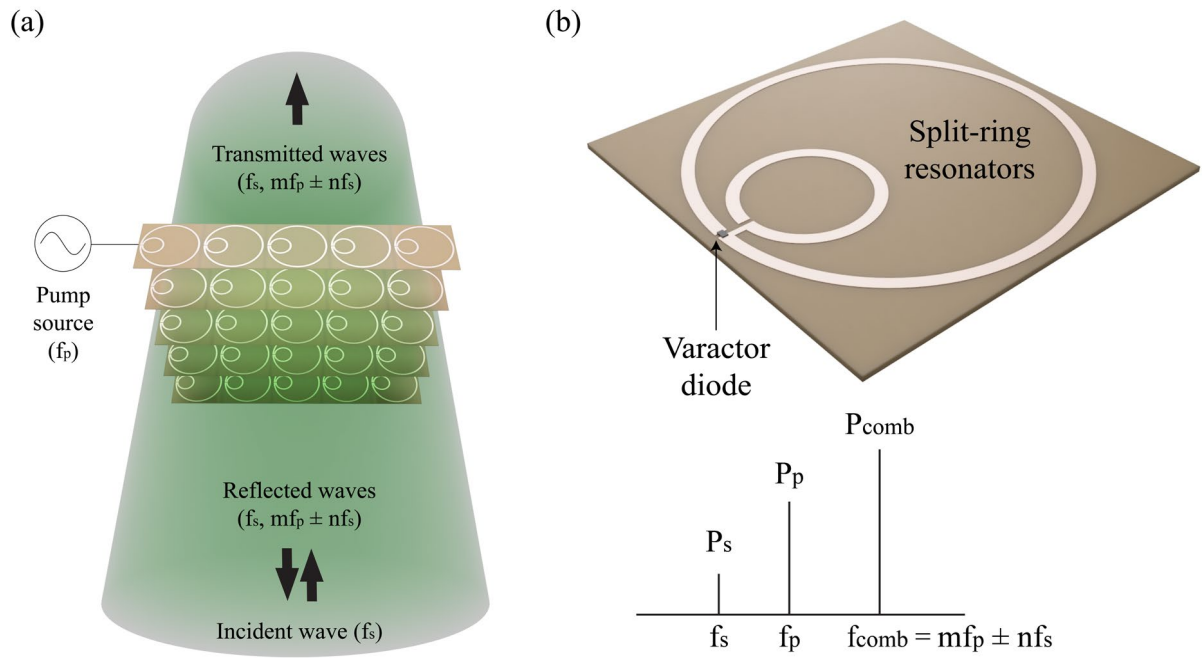

FIG 1. (a) Schematic representation of parametric amplification and frequency conversion by a metasurface. An electromagnetic wave at frequency $f_s$ is incident on a parametric metasurface modulated by a pump source at frequency $f_p$. (b) The meta-atom structure consisting of two connected split-ring resonators with an embedded varactor diode. The meta-atom utilizes parametric interaction of the incident wave at $f_s$ and the pump source at $f_p$, achieving up-conversion to the combination frequencies $f_{comb}$.

This is achieved by using varactor diodes with parametrically modulated capacitance, which connect two resonators in each meta-atom. We assume that the capacitance oscillates periodically with the pump frequency:

$$C(t) = C_0[1 + \beta \cos(\omega_p t + \varphi_p)], \qquad (1)$$

where $C_0$ is the mean value of the capacitance, $\beta = \frac{\Delta C}{C_0}$ is the modulation coefficient, $\Delta C$ is the amplitude of the capacitance modulation, $\omega_p = 2\pi f_p$ is the pump angular frequency, $\varphi_p$ is the phase of the pump wave.

Modulation of the varactor is achieved by the pump source with a voltage given by

$$u_p(t) = U_0 + U_p \cos(\omega_p t + \varphi_p), \tag{2}$$

where $U_0$ is the constant bias voltage, $U_p$ is the pump amplitude.

The maximum possible conversion efficiency for the combination frequency $f_{comb}$ is given by the Manley-Rowe relations[43]:

$$\begin{cases} \frac{P_s}{f_s} + \frac{mP_{comb}}{f_{comb}} = 0 \\ \frac{P_p}{f_p} + \frac{nP_{comb}}{f_{comb}} = 0 \end{cases} \tag{3}$$

For $m = n = 1$ the Manley-Rowe relations can be simplified as follows:

$$\begin{cases} \frac{P_s}{f_s} + \frac{P_{sum}}{f_{sum}} = 0 \\ \frac{P_p}{f_p} + \frac{P_{sum}}{f_{sum}} = 0 \end{cases} \tag{4}$$

In this case, the maximum gain is determined by the frequency ratio:

$$K_P = \frac{P_{sum}}{P_s} = \frac{f_{sum}}{f_s}, \tag{5}$$

where $P_{sum} = P_s + P_p$, $P_s$ is the power of the incident wave, $P_p$ is the power of the pump source. Both the incident wave and the pump source contribute to the generated power at the sum-frequency.

Figure 2(a) depicts the meta-atom comprising two connected copper split-ring resonators positioned on the Rogers RO4003C substrate, with the embedded varactor diode MGV100-08 shown as a black square. The selection of this varactor diode is due to its wide capacitance adjustment range, with a very low minimum capacitance, enabling the necessary tuning of the system's resonant frequencies. The varactor diode is modulated by the pump source through the following filtering circuit: the passband LC filter at $f_p$ serially connected to the AC source and the lowpass RL filter with a cutoff frequency of 16 MHz connected to the DC source. The DC source is used to set the operating bias voltage ($U_0$) on the capacitance-voltage characteristic of the varactor diode. To avoid additional losses in the pump circuit, we keep the following relation when changing the pump frequency:

$$L_{AC} = \frac{1}{(2\pi f_p)^2 C_{AC}}, \qquad (6)$$

where $L_{AC}$ is the inductance and $C_{AC} = 0.1$ pF is the capacitance used in the bandpass LC-filter.

The split-ring resonators have the following parameters shown in Fig. 2(a): $R_1 = 34$ mm, $R_2 = 14$ mm, $w = 2$ mm, $g_1 = 0.5$ mm, $g_2 = 1.25$ mm, $d = 0.5$ mm. The thickness of the split-ring resonators is 35 µm, the substrate thickness is 508 µm, and the unit cell size is 70 mm x 70 mm.

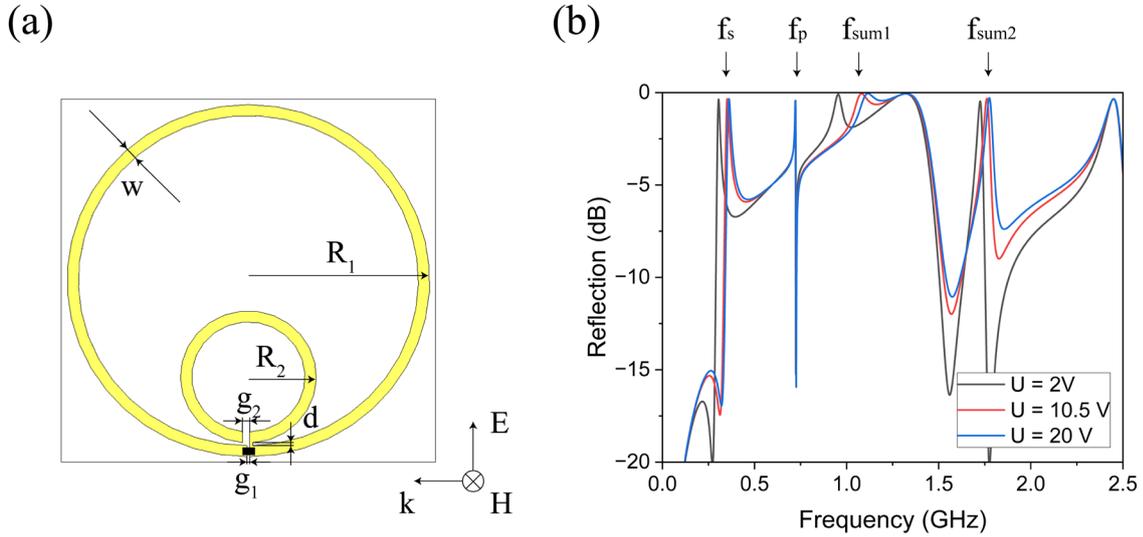

FIG 2. (a) The meta-atom design consisting of two copper split-ring resonators placed on the substrate and the embedded varactor diode shown as a black square. (b) The reflection characteristics of the proposed metasurface for three DC voltage values applied to the varactor diodes.

The metasurface is an array of meta-atoms shown in Fig. 1(a), supporting multiple resonances. We conduct full-wave numerical simulations of the periodic structure using the finite-element method and transfer the 3D modeling results to the Schematic simulator in CST Studio to model the nonlinear lumped element circuit, which includes the varactor diode and the biasing network. Fig. 2(b) illustrates the reflection characteristics of the proposed metasurface for three different DC voltage values applied to the varactor diodes when the incident plane wave with one of the polarizations propagates along one of the substrate axes. In this work we utilize the resonances at $f_s$, $f_{sum1}$ and $f_{sum2}$ shown in Fig. 2(b). The resonance at $f_p$ is not related to the split-ring resonators and their structure, it appears due to a bandpass LC-filter connecting the pump source and varactors in meta-atoms.

The proposed parametric metasurface operates in the degenerate regime when

$$f_p = 2f_s, \tag{7}$$

$$f_{sum1} = f_p + f_s = 3f_s, \tag{8}$$

$$f_{sum2} = 2f_p + f_s = f_p + 3f_s = 5f_s. \tag{9}$$

We found that using a pump at twice the signal frequency and tuning the SRRs to resonate near integer multiples of the signal frequency enables the generation of higher-power up-converted electromagnetic waves compared to using frequencies that deviate significantly from these multiples. This occurs because of a cascaded (multi-stage) amplification and up-conversion process, in which the pump source amplifies the signal and up-converts it to the sum-frequencies. As a result, it is possible to exceed the maximum conversion efficiency determined by the frequency ratio in Eq. 5. However, the frequency ratio still dictates the *maximum* attainable gain relative to the amplified power of the incident wave:

$$K_P = \kappa \frac{f_{sum}}{f_s} = \kappa \frac{P_{sum}}{P'_s} = \frac{P_{sum}}{P_s}, \tag{10}$$

where $\kappa = P'_s/P_s$ is the amplification coefficient for the incident wave at $f_s$. Since the frequency ratio remains unchanged, but the signal is amplified, a higher power level can be achieved at the sum-frequency. The cascaded process begins with parametric amplification of currents at the signal frequency, induced by the incident electromagnetic waves in the split-ring resonators[22,37]. These currents are further mixed with the currents excited at the pump frequency, producing currents at the sum-frequencies, which then radiate up-converted electromagnetic waves.

We utilize the Spectral Lines task and the varactor SPICE model within the Schematic module of CST Studio to simulate the nonlinear circuit in the frequency domain. This approach leverages Harmonic Balance (HB) analysis, allowing us to investigate the interaction between the pump and signal within the designed circuit. The method accounts for frequency mixing up to a defined harmonic limit. To accurately capture the mixing effects, we consider 20 harmonics for the signal source and 10 harmonics for the pump source.

First, we determine the frequency at which the maximum power of the radiated waves occurs in the degenerate mode. Fig. 3(a-c) depict the dependence of the power of the amplified electromagnetic waves at $f_s$, $f_{sum1}$ and $f_{sum2}$ on the frequency for $U_{DC} = 10.5$ V and $U_p = 8$ V, when both $f_s$ and $f_p$ change under the condition of phase-locking in the degenerate regime. The blue and red lines represent the power radiated in different directions.

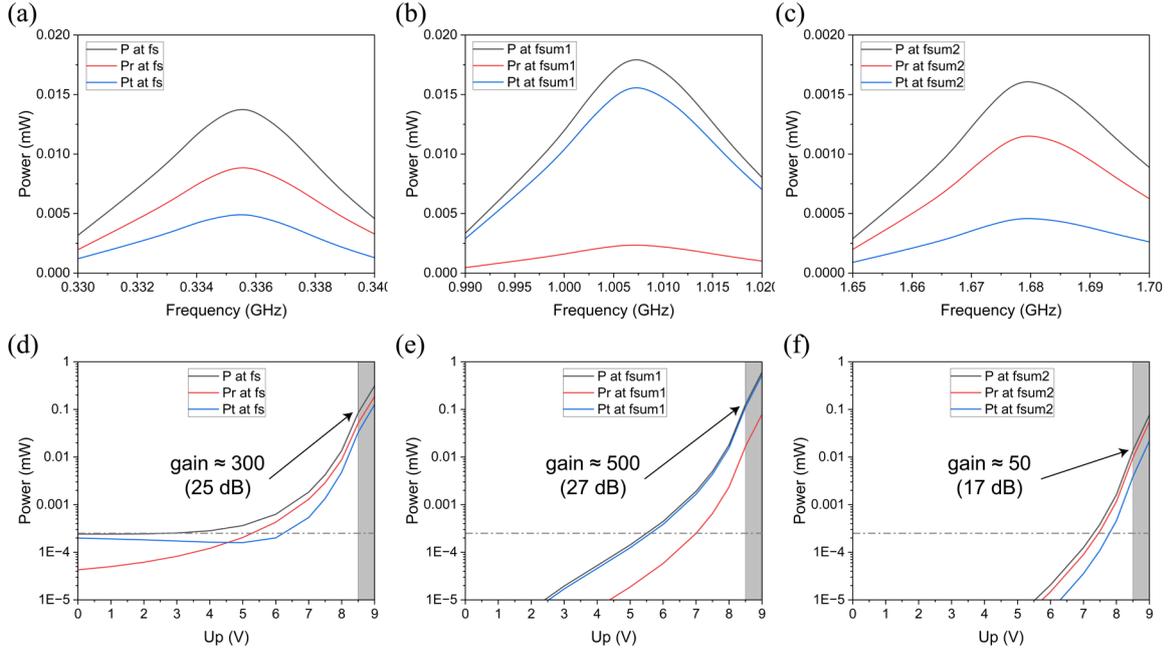

FIG 3. (a-c) The dependence of the power of the amplified electromagnetic waves at $f_s$, $f_{sum1}$ and $f_{sum2}$ on the frequency for $U_{DC} = 10.5$ V and $U_p = 8$ V under the condition of phase-locking in the degenerate regime. (d-f) The dependence of the power of the radiated electromagnetic waves at $f_s = 0.336$ GHz, $f_{sum1} = 1.008$ GHz and $f_{sum2} = 1.68$ GHz on the pump amplitude ($U_p$) at $f_p = 0.672$ GHz for $U_{DC} = 10.5$ V under the condition of phase-locking in the degenerate regime (logarithmic scale). The blue and red lines indicate the power radiated to different directions. The horizontal dotted line shows the incident electromagnetic wave power level (0.25 µW). The area when the varactor model becomes inaccurate (for $U_p > 8.5$ V) is shaded.

Figures 3(d-f) show the dependence of the power of the amplified electromagnetic waves at $f_s = 0.336$ GHz, $f_{sum1} = 1.008$ GHz and $f_{sum2} = 1.68$ GHz on the pump amplitude at $f_p = 0.672$ GHz for $U_{DC} = 10.5$ V under the condition of phase-locking in the degenerate regime. The horizontal dotted line shows the incident electromagnetic wave power level (0.25 µW). The area when the varactor model becomes inaccurate according to the varactor manufacturer ($U_p > 8.5$ V) is shaded. We predict amplification of more than 300 times (25 dB) at $f_s$, with the power of the radiated electromagnetic waves at $f_{sum1}$ being more than 500 times (27 dB) and at $f_{sum2}$ being more than 50 times (17 dB) higher than the power of the incident

wave in the degenerate regime. The radiated power at $f_s$, $f_{sum1}$ and $f_{sum2}$ depends on the pump phase in the degenerate regime as it is shown in Fig. 4 for $U_{DC} = 10.5$ V and $U_p = 8$ V.

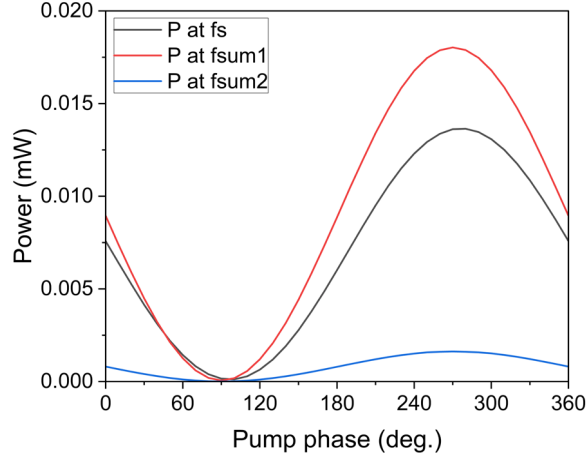

FIG 4. The dependence of the radiated power at $f_s = 0.336$ GHz, $f_{sum1} = 1.008$ GHz and $f_{sum2} = 1.68$ GHz on the pump phase at $f_p = 0.672$ GHz for $U_{DC} = 10.5$ V and $U_p = 8$ V in the degenerate regime.

Next, we analyze the operation of the proposed parametric metasurface in the non-degenerate regime when the frequency ratio given in Eq. 7-9 is not satisfied. This process is less efficient due to the additional generation of electromagnetic waves at idler frequencies $f_i$, $f_{sum1i}$ and $f_{sum2i}$. The non-degenerate regime is characterized by the following frequency ratio: $f_i = f_p - f_s, f_{sum1} = f_p + f_s, f_{sum1i} = f_p + f_i, f_{sum2} = 2f_p + f_s, f_{sum2i} = 2f_p + f_i$.

In this section, we analyze two non-degenerate regimes: a) operating near the degenerate regime in Eq.7-9; b) operating at frequencies far from those of the degenerate regime. This allows us to compare two different non-degenerate regimes and demonstrate that higher conversion efficiency can be achieved at frequencies close to the degenerate case, while also exceeding the converted power limits imposed by the Manley-Rowe relations in Eq. 5.

Figures 5(a-c) show the dependence of the power of the radiated electromagnetic waves at $f_s$, $f_{sum1}$, $f_{sum2}$ (solid lines), $f_i$, $f_{sum1i}$ and $f_{sum2i}$ (dotted lines) on the frequency in the non-degenerate regime for $U_{DC} = 10.5$ V, $U_p = 8.5$ V and $f_s = 0.336$ GHz. The horizontal dotted line shows the incident electromagnetic wave power level (0.25 µW). Although the degenerate regime (corresponds to the narrow peaks when $f_p = 0.672$ GHz) allows for achieving significantly higher level of output power at both sum-frequencies and the signal frequency, its practical realization is challenging due to its extremely narrow band. In addition to phase locking, fine tuning of the pump and signal sources is necessary. While the power of the radiated electromagnetic waves in the non-degenerate regime is lower, it is still possible to achieve a level at both sum-frequencies that exceeds the incident wave power.

Next, we analyze and compare two non-degenerate regimes with pump frequencies of $f_p = 0.671$ GHz and $f_p = 0.74$ GHz, respectively. These scenarios are labeled as 1 and 2 in Figure 5(a-c) for clarity.

Figures 5(a-c) show that the non-degenerate regime operating near the degenerate resonance frequency ratio allows for achieving significant level of the radiated power at $f_s$, $f_{sum1}$, $f_{sum2}$ (solid lines), $f_i$, $f_{sum1i}$ and $f_{sum2i}$ (dotted lines). The 3-dB bandwidth of the up-conversion at $f_{sum1}$ (close to 1.008 GHz) and at $f_{sum2}$ (close to 1.68 GHz) is around 3 and 6 MHz respectively.

Figures 5(d-f) show the dependence of the power of the radiated electromagnetic waves at $f_s = 0.336$ GHz, $f_i = 0.335$ GHz, $f_{sum1} = 1.007$ GHz, $f_{sum1i} = 1.006$ GHz, $f_{sum2} = 1.678$ GHz and $f_{sum2i} = 1.677$ GHz on the pump amplitude at $f_p = 0.671$ GHz for $U_{DC} = 10.5$ V in this non-degenerate regime. The area when the varactor model becomes inaccurate (for $U_p = 8.5$ V) is shaded. We predict amplification of more than 70 times (18.5 dB) at $f_s$, with the power of the radiated electromagnetic waves at $f_{sum1}$ being more than 120 times (21 dB) and at $f_{sum2}$ being more than 10 times (10 dB) higher than the power of the incident wave in the non-degenerate regime within the limits of the varactor model.

Figure 5(b) also shows that the power of the radiated electromagnetic waves around $f_{sum1} = 1.076$ GHz is higher than the power of the incident wave. This case corresponds to the sum-frequency value close to the reflection maximum when U = 10.5 V shown in Fig. 2(b) (red curve). We expect that the precise tuning of the resonant frequencies of the split-ring resonators will result in the absence of this additional peak in the up-conversion spectra and allow us to achieve higher power levels of radiated electromagnetic waves at frequencies close to the degenerate resonance frequency ratio in Eq.7-9. However, further optimization of the parametric system lies beyond the scope of this article and our current resources. Additionally, this slight detuning allows us to use the same system for comparing two cases of the non-degenerate regimes as discussed above.

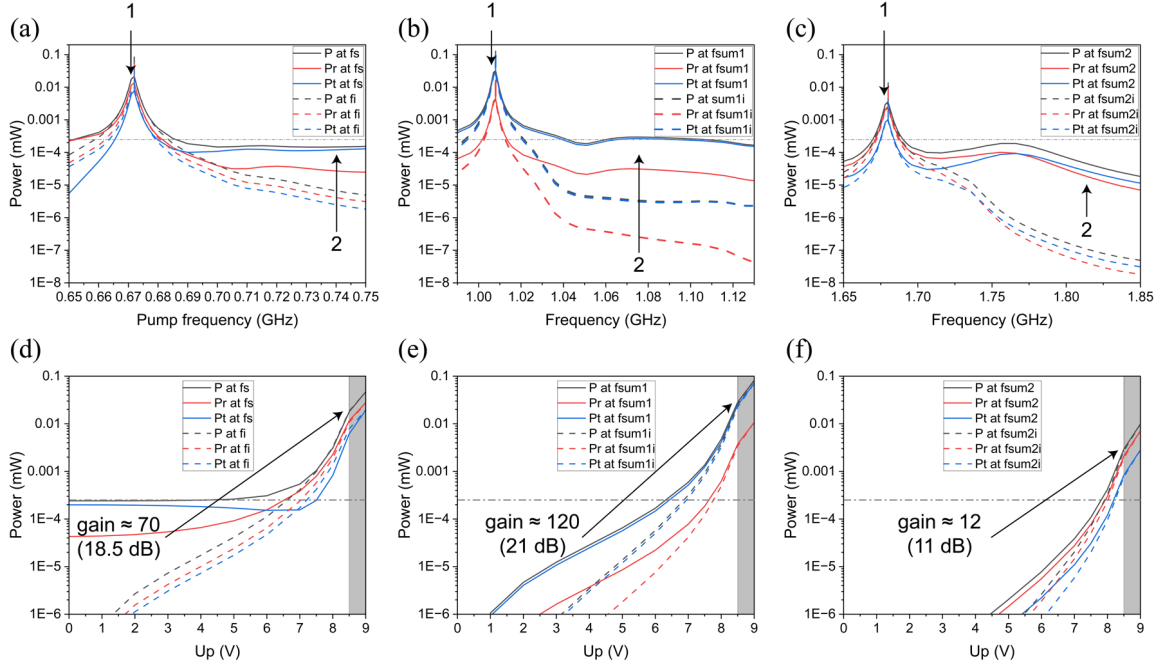

FIG 5. (a-c) The dependence of the power of the radiated electromagnetic waves at $f_s$, $f_{sum1}$, $f_{sum2}$ (solid lines), $f_i$, $f_{sum1i}$ and $f_{sum2i}$ (dotted lines) on the frequency in the non-degenerate regime for $U_{DC}$ = 10.5 V and $U_p$ = 8.5 V when $f_s$ = 0.336 GHz and $f_p$ changes (logarithmic scale). The blue and red lines indicate the transmitted and reflected power (radiated by the metasurface in different directions), while the horizontal dotted line shows the incident electromagnetic wave power level (0.25 µW). (d-f) The dependence of the power of the radiated electromagnetic waves at $f_s$ = 0.336 GHz, $f_i$ = 0.335 GHz, $f_{sum1}$ = 1.007 GHz, $f_{sum1i}$ = 1.006 GHz, $f_{sum2}$ = 1.678 GHz and $f_{sum2i}$ = 1.677 GHz on the pump amplitude ($U_p$) at $f_p$ = 0.671 GHz for $U_{DC}$ = 10.5 V in the non-degenerate regime (logarithmic scale). The area when the varactor model becomes inaccurate (for $U_p$ > 8.5 V) is shaded.

Next, we analyze the non-degenerate regime when $f_p$ = 0.74 GHz and $f_s$ = 0.336 GHz. In this case, there is no amplification of the electromagnetic waves at $f_s$, and the power of the incident waves is utilized for up-conversion. Fig. 6 illustrates a decrease in the radiated power at $f_s$ (a) and an increase in the radiated power at both sum-frequencies (b) as the pump amplitude increases, with the power at $f_{sum1}$ exceeding the incident power when $U_p$ > 7.5 V. This contrasts with the previously discussed non-degenerate regime operating near the degenerate resonance frequency regime, when we observe the exponential growth of the radiated power at $f_s$, $f_{sum1}$ and $f_{sum2}$ with increasing the modulation amplitude due to the cascaded amplification process. The dependence of the radiated power at $f_{sum1}$ on the pump amplitude shows a decrease in its growth rate when $U_p$ approaches 8.5 V.

The maximum radiated power at $f_{sum1}$ in the non-degenerate regime operating at frequencies far from the degenerate regime is approximately 0.3 µW for $U_{DC}$ = 10.5 V, $U_p$ = 8.5 V, $f_p$ = 0.74 GHz and $f_s$ = 0.336 GHz. This results in a conversion efficiency of 1.2 (not exceeding 1 dB), which is significantly lower compared to the non-degenerate regime operating near the degenerate resonance frequency ratio with the same pump amplitude, where the conversion efficiency is approximately 120 (exceeding 20 dB) for $U_{DC} = 10.5$ V, $U_p = 8.5$ V, $f_p = 0.671$ GHz and $f_s = 0.336$ GHz due to the cascaded amplification and up-conversion process.

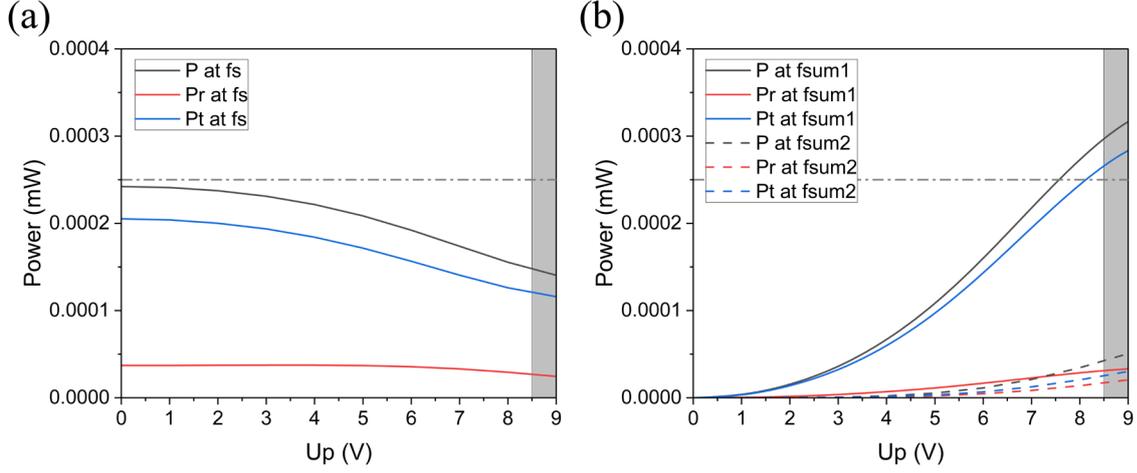

FIG 6. The dependence of the power of the radiated electromagnetic waves at $f_s = 0.336$ GHz (a), $f_{sum1} = 1.076$ GHz and $f_{sum2} = 1.816$ GHz (b) on the pump amplitude ($U_p$) at $f_p = 0.74$ GHz for $U_{DC} = 10.5$ V in the non-degenerate regime. The blue and red lines indicate the power radiated to different directions. The area when the varactor model becomes inaccurate (for $U_p > 8.5$ V) is shaded.

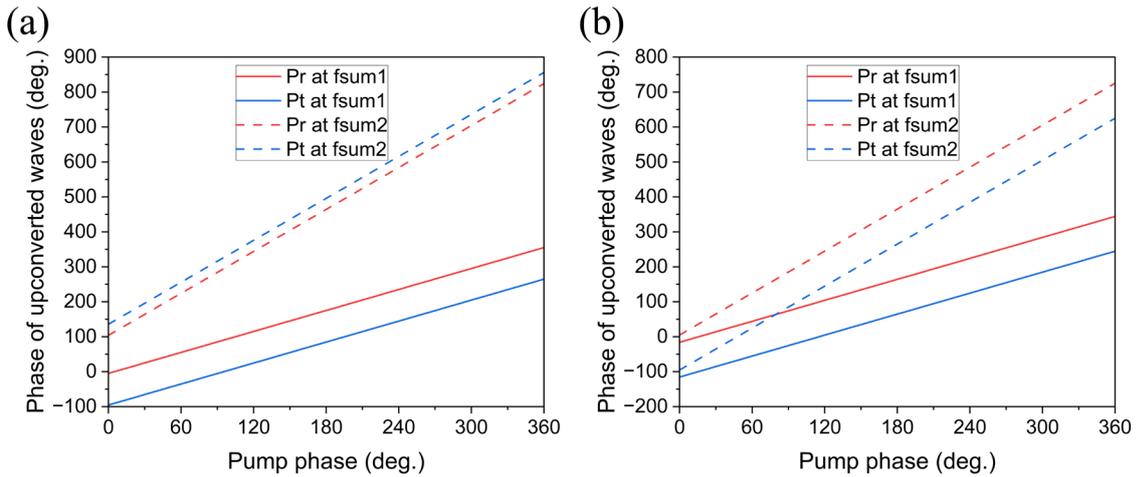

FIG 7. The dependence of the radiated electromagnetic wave phase at both sum-frequencies $f_{sum1}$ and $f_{sum2}$ on the pump phase in two non-degenerate regimes: when $f_p = 0.671$ (a) and $f_p = 0.74$ GHz (b).

Additionally, we study the dependence of the phase of the radiated electromagnetic waves at both sum-frequencies on the pump phase. We observe the linear relationship between the phases in the non-degenerate regime when $f_p$ = 0.671 and 0.74 GHz, as depicted in Fig. 7. This suggests that by choosing an appropriate pump we can control the phase of the upconverted waves. Since the phase modulation can be made non-uniform across the metasurface, the proposed approach can be utilized not only for amplifying and up-converting electromagnetic waves but also for spatial control of the converted waves. The phase gradient across the array of the meta-atoms opens possibilities for beam steering and other forms of control[16,26] useful in communication applications.

The case of traditional parametric amplification, when $f_p = 2f_s$ is well known and several metasurface implementations have recently been proposed[22,37]. In contrast, our study demonstrates the ability to convert a signal with amplification to a combination frequency above the pump frequency.

According to the theory of parametric amplification[43], the maximum gain is proportional to the ratio of the combination frequency to the incident wave frequency. However, we found that in the case of the pump and sum-frequencies near integer multiples of the signal frequency, the power of the radiated waves at the sum-frequencies can be increased compared to the case of the arbitrary choice of the pump frequency due to the cascaded amplification and up-conversion process. This allows us to overcome the converted power limits imposed by the Manley-Rowe relations. However, the pump frequency should be close to twice the signal frequency, which necessitates up-conversion to a sum-frequency approximately three times higher than the signal frequency. Additionally, the proposed parametric system has several drawbacks, including beatings in the non-degenerate regime when operating at frequencies near the degenerate frequency ratio, as well as the requirement to strictly maintain the frequency ratio and a fixed phase difference between the incident wave and the pump in the degenerate regime.

The frequency conversion ratio can be enhanced by employing modulation at several pump frequencies, optimizing resonator design, and stacking multiple metasurface layers. One of the possible ways for increasing the power and bandwidth of the up-converted electromagnetic waves in the non-degenerate regime is using dual-tone modulation techniques with two pump sources at frequencies close to the degenerate case $f_p = 2f_s$. This will also allow us to utilize the power radiated at idler frequencies for further improvement of the up-conversion efficiency. Another possible way for improving the efficiency of conversion is appropriate meta-atom design not radiating power at $f_p$.

Furthermore, as demonstrated, amplified conversion can be achieved at multiple sum-frequencies using a single signal frequency of the incident electromagnetic wave. This capability is particularly useful for designing frequency synthesizer systems in communication networks, enabling the utilization of multiple radio channels. Due to the widespread application of parametric amplification in quantum computing readouts and receivers in radio astronomy and deep space communication, the proposed metasurfaces could significantly advance current technologies in this domain[44–46]. The obtained results can also be applied in the design of time crystals[41,42].

The proposed parametric metasurfaces enable up-conversion with amplification of free-space electromagnetic waves. Our modeling results highlight the potential of using a pump at twice the signal frequency and resonances near integer multiples of the signal frequency for utilizing a cascaded process, which allows to overcome the limitations imposed by the Manley-Rowe relations. The integration of up-conversion and amplification within a single metasurface is highly promising for communication systems. These methods can be applied across the radio, microwave, and subterahertz ranges, currently constrained by the properties of modern varactor diodes, with the potential to extend to higher frequencies using ultrafast modulation techniques.

This research was supported by the Australian Research Council Centre of Excellence for Transformative Meta-Optical Systems (Project ID CE200100010). The authors thank David Powell for useful discussions.

Data underlying the results presented in this paper are not publicly available at this time but may be obtained from the authors upon reasonable request.

The authors declare no conflicts of interest.

# REFERENCES

[1] A. Li, S. Singh, and D. Sievenpiper, "Metasurfaces and their applications," Nanophotonics **7**(6), 989–1011 (2018).
[2] D.N. Neshev, and A.E. Miroshnichenko, "Enabling smart vision with metasurfaces," Nat. Photon. **17**(1), 26–35 (2023).
[3] W.T. Chen, A.Y. Zhu, and F. Capasso, "Flat optics with dispersion-engineered metasurfaces," Nat Rev Mater **5**(8), 604–620 (2020).
[4] Y. He, B. Song, and J. Tang, "Optical metalenses: fundamentals, dispersion manipulation, and applications," Front. Optoelectron. **15**(1), 24 (2022).
[5] T. Badloe, J. Lee, J. Seong, and J. Rho, "Tunable Metasurfaces: The Path to Fully Active Nanophotonics," Adv Photo Res **2**(9), 2000205 (2021).
[6] O.A.M. Abdelraouf, Z. Wang, H. Liu, Z. Dong, Q. Wang, M. Ye, X.R. Wang, Q.J. Wang, and H. Liu, "Recent Advances in Tunable Metasurfaces: Materials, Design, and Applications," ACS Nano **16**(9), 13339–13369 (2022).
[7] J. Kim, J. Seong, Y. Yang, S.-W. Moon, T. Badloe, and J. Rho, "Tunable metasurfaces towards versatile metalenses and metaholograms: a review," Adv. Photon. **4**(02), (2022).
[8] J. Yang, S. Gurung, S. Bej, P. Ni, and H.W. Howard Lee, "Active optical metasurfaces: comprehensive review on physics, mechanisms, and prospective applications," Rep. Prog. Phys. **85**(3), 036101 (2022).
[9] T. Gu, H.J. Kim, C. Rivero-Baleine, and J. Hu, "Reconfigurable metasurfaces towards commercial success," Nat. Photon. **17**(1), 48–58 (2023).
[10] A.M. Shaltout, V.M. Shalaev, and M.L. Brongersma, "Spatiotemporal light control with active metasurfaces," Science **364**(6441), eaat3100 (2019).
[11] C. Caloz, and Z.-L. Deck-Leger, "Spacetime Metamaterials—Part I: General Concepts," IEEE Trans. Antennas Propagat. **68**(3), 1569–1582 (2020).
[12] C. Caloz, and Z.-L. Deck-Leger, "Spacetime Metamaterials—Part II: Theory and Applications," IEEE Trans. Antennas Propagat. **68**(3), 1583–1598 (2020).
[13] S. Taravati, and G.V. Eleftheriades, "Microwave Space-Time-Modulated Metasurfaces," ACS Photonics **9**(2), 305–318 (2022).
[14] E. Galiffi, R. Tirole, S. Yin, H. Li, S. Vezzoli, P.A. Huidobro, M.G. Silveirinha, R. Sapienza, A. Alù, and J.B. Pendry, "Photonics of time-varying media," Adv. Photon. **4**(01), (2022).
[15] N. Engheta, "Four-dimensional optics using time-varying metamaterials," Science **379**(6638), 1190–1191 (2023).
[16] M. Liu, D.A. Powell, Y. Zarate, and I.V. Shadrivov, "Huygens' Metadevices for Parametric Waves," Phys. Rev. X **8**(3), 031077 (2018).
[17] V. Asadchy, A.G. Lamprianidis, G. Ptitcyn, M. Albooyeh, Rituraj, T. Karamanos, R. Alaee, S.A. Tretyakov, C. Rockstuhl, and S. Fan, "Parametric Mie Resonances and Directional Amplification in Time-Modulated Scatterers," Phys. Rev. Applied **18**(5), 054065 (2022).
[18] Z. Hayran, and F. Monticone, "Using Time-Varying Systems to Challenge Fundamental Limitations in Electromagnetics: Overview and summary of applications," IEEE Antennas Propag. Mag. **65**(4), 29–38 (2023).
[19] A. Shaltout, A. Kildishev, and V. Shalaev, "Time-varying metasurfaces and Lorentz non-reciprocity," Opt. Mater. Express **5**(11), 2459 (2015).
[20] Y. Hadad, D.L. Sounas, and A. Alu, "Space-time gradient metasurfaces," Phys. Rev. B **92**(10), 100304 (2015).
[21] T. Liu, J.-Y. Ou, K.F. MacDonald, and N.I. Zheludev, "Photonic metamaterial analogue of a continuous time crystal," Nat. Phys. **19**(7), 986–991 (2023).

[22] X. Wang, M.S. Mirmoosa, V.S. Asadchy, C. Rockstuhl, S. Fan, and S.A. Tretyakov, "Metasurface-based realization of photonic time crystals," Sci. Adv. **9**(14), eadg7541 (2023).

[23] R. Tirole, S. Vezzoli, E. Galiffi, I. Robertson, D. Maurice, B. Tilmann, S.A. Maier, J.B. Pendry, and R. Sapienza, "Double-slit time diffraction at optical frequencies," Nat. Phys. **19**(7), 999–1002 (2023).

[24] L. Zhang, X.Q. Chen, S. Liu, Q. Zhang, J. Zhao, J.Y. Dai, G.D. Bai, X. Wan, Q. Cheng, G. Castaldi, V. Galdi, and T.J. Cui, "Space-time-coding digital metasurfaces," Nat Commun **9**(1), 4334 (2018).

[25] S. Taravati, and G.V. Eleftheriades, "Full-Duplex Nonreciprocal Beam Steering by Time-Modulated Phase-Gradient Metasurfaces," Phys. Rev. Applied **14**(1), 014027 (2020).

[26] K. Lee, J. Son, J. Park, B. Kang, W. Jeon, F. Rotermund, and B. Min, "Linear frequency conversion via sudden merging of meta-atoms in time-variant metasurfaces," Nature Photon **12**(12), 765–773 (2018).

[27] M. Liu, A.B. Kozyrev, and I.V. Shadrivov, "Time-varying Metasurfaces for Broadband Spectral Camouflage," Phys. Rev. Applied **12**(5), 054052 (2019).

[28] Z. Wu, and A. Grbic, "Serrodyne Frequency Translation Using Time-Modulated Metasurfaces," IEEE Trans. Antennas Propagat. **68**(3), 1599–1606 (2020).

[29] S. Taravati, and G.V. Eleftheriades, "Pure and Linear Frequency-Conversion Temporal Metasurface," Phys. Rev. Applied **15**(6), 064011 (2021).

[30] L. Wang, H. Shi, G. Peng, J. Yi, L. Dong, A. Zhang, and Z. Xu, "A Time-Modulated Transparent Nonlinear Active Metasurface for Spatial Frequency Mixing," Materials **15**(3), 873 (2022).

[31] Y. Jiang, S. Duan, J. Wu, H. Qiu, C. Zhang, K. Fan, H. Wang, B. Jin, J. Chen, and P. Wu, "Active terahertz frequency conversion on silicon-based time-varying metasurface," Applied Physics Letters **123**(16), 161702 (2023).

[32] R. Tirole, S. Vezzoli, D. Saxena, S. Yang, T.V. Raziman, E. Galiffi, S.A. Maier, J.B. Pendry, and R. Sapienza, "Second harmonic generation at a time-varying interface," Nat Commun **15**(1), 7752 (2024).

[33] E. Galiffi, P.A. Huidobro, and J.B. Pendry, "Broadband Nonreciprocal Amplification in Luminal Metamaterials," PHYSICAL REVIEW LETTERS, 6 (2019).

[34] Z. Seyedrezaei, B. Rejaei, and M. Memarian, "Frequency conversion and parametric amplification using a virtually rotating metasurface," Opt. Express **28**(5), 6378 (2020).

[35] S. Taravati, and G.V. Eleftheriades, "Full-duplex reflective beamsteering metasurface featuring magnetless nonreciprocal amplification," Nat Commun **12**(1), 4414 (2021).

[36] X. Wang, J. Han, S. Tian, D. Xia, L. Li, and T.J. Cui, "Amplification and Manipulation of Nonlinear Electromagnetic Waves and Enhanced Nonreciprocity using Transmissive Space-Time-Coding Metasurface," Advanced Science **9**(11), 2105960 (2022).

[37] F.V. Kovalev, and I.V. Shadrivov, "Parametric metasurfaces for electromagnetic wave amplification," Opt. Mater. Express **14**(2), 494 (2024).

[38] Z. Seyedrezaei, B. Rejaei, and M. Memarian, "Synthetic rotational Doppler shift on transmission lines and it's microwave applications," Sci Rep **14**(1), 21303 (2024).

[39] X. Wen, X. Zhu, A. Fan, W.Y. Tam, J. Zhu, H.W. Wu, F. Lemoult, M. Fink, and J. Li, "Unidirectional amplification with acoustic non-Hermitian space−time varying metamaterial," Commun Phys **5**(1), 18 (2022).

[40] A.G. Löhr, M.Y. Ivanov, and M.A. Khokhlova, "Controlled compression, amplification and frequency up-conversion of optical pulses by media with time-dependent refractive index," Nanophotonics **12**(14), 2921–2928 (2023).

[41] S. Saha, O. Segal, C. Fruhling, E. Lustig, M. Segev, A. Boltasseva, and V.M. Shalaev, "Photonic time crystals: a materials perspective [Invited]," Opt. Express **31**(5), 8267 (2023).

[42] A. Boltasseva, V.M. Shalaev, and M. Segev, "Photonic time crystals: from fundamental insights to novel applications: opinion," Opt. Mater. Express **14**(3), 592 (2024).

[43] J. Manley, and H. Rowe, "Some General Properties of Nonlinear Elements-Part I. General Energy Relations," Proc. IRE **44**(7), 904–913 (1956).

[44] J. Aumentado, "Superconducting Parametric Amplifiers: The State of the Art in Josephson Parametric Amplifiers," IEEE Microwave **21**(8), 45–59 (2020).

[45] L. Fasolo, A. Greco, and E. Enrico, "Superconducting Josephson-Based Metamaterials for Quantum-Limited Parametric Amplification: A Review," in *Advances in Condensed-Matter and Materials Physics - Rudimentary Research to Topical Technology*, edited by J. Thirumalai and S. Ivanovich Pokutnyi, (IntechOpen, 2020).

[46] D.J. Parker, M. Savytskyi, W. Vine, A. Laucht, T. Duty, A. Morello, A.L. Grimsmo, and J.J. Pla, "Degenerate Parametric Amplification via Three-Wave Mixing Using Kinetic Inductance," Phys. Rev. Applied **17**(3), 034064 (2022).

[47] S. I. Baskakov, Signals and Circuits (Mir, 1986).